\pdfoutput=1
\documentclass[11pt]{article}
\usepackage{cite}
\usepackage[]{acl}

\usepackage{times}
\usepackage{latexsym}

\usepackage[T1]{fontenc}
\usepackage[utf8]{inputenc}

\usepackage{microtype}

\usepackage{color}
\usepackage{amsmath}
\usepackage{graphicx}
\usepackage{arydshln} % dashline
\usepackage{multirow}
\definecolor{c1}{HTML}{95bddc}
\definecolor{c2}{HTML}{c2d1e5}
\definecolor{c3}{HTML}{fe793d}
\definecolor{c4}{HTML}{fb4c1f}
\definecolor{c5}{HTML}{b71a3b}
\definecolor{c6}{HTML}{7e0f12}
\definecolor{c7}{HTML}{E85642}
\definecolor{c8}{HTML}{C00000}
\definecolor{c9}{HTML}{ff2d51}
\definecolor{blue}{HTML}{339ff4}
\definecolor{green}{HTML}{3ca057}
\definecolor{blue_deep}{HTML}{517293}
\title{Enhancing Empathetic Response Generation by Augmenting LLMs with Small-scale Empathetic Models}

\author{
Zhou Yang$^{1,2}$, Zhaochun Ren$^{3}$, Yufeng Wang$^{1,2}$, \textbf{Shizhong Peng$^{1,2}$}, \\
\textbf{Haizhou Sun$^4$}, \textbf{Xiaofei Zhu$^{5}$},  
\textbf{Xiangwen Liao}$^{1,2}$\thanks{\hspace{1mm} Corresponding author.}\\
\small $^1$College of Computer and Data Science, Fuzhou University; 
$^2$Digital Fujian Institute of Financial Big Data,
 Fuzhou, China \\ 
\small $^3$Leiden University, Leiden, The Netherlands;
\small $^4$H. Sun is with SmartMore \\
\small $^5$College of Computer Science and Technology, Chongqing University of Technology, Chongqing, China\\ 
\small  \texttt{\{200310007, 211027083, 102102153, liaoxw\}@fzu.edu.cn} \\
\small  \texttt{z.ren@liacs.leidenuniv.nl} \hspace{0.1cm} \texttt{zxf@cqut.edu.cn}\\
}

\begin{document}
\maketitle

\begin{abstract}
Empathetic response generation is increasingly significant in AI, necessitating nuanced emotional and cognitive understanding coupled with articulate response expression.
Current large language models (LLMs) excel in response expression;
however, they lack the ability to deeply understand emotional and cognitive nuances, particularly in pinpointing fine-grained emotions and their triggers.
Conversely, small-scale empathetic models (SEMs) offer strength in fine-grained emotion detection and 
detailed emotion cause identification.
To harness the complementary strengths of both LLMs and SEMs, we introduce a Hybrid Empathetic Framework (HEF). 
HEF regards SEMs as flexible plugins to improve LLM's nuanced emotional and cognitive understanding.
Regarding emotional understanding, HEF implements a two-stage emotion prediction strategy, encouraging LLMs to prioritize primary emotions emphasized by SEMs, followed by other categories, substantially alleviates the difficulties for LLMs in fine-grained emotion detection.
Regarding cognitive understanding, HEF employs an emotion cause perception strategy, prompting LLMs to focus on crucial emotion-eliciting words identified by SEMs, thus boosting LLMs' capabilities in identifying emotion causes.
This collaborative approach enables LLMs to discern emotions more precisely and formulate empathetic responses. We validate HEF on the Empathetic-Dialogue dataset, and the findings indicate that our framework enhances the refined understanding of LLMs and their ability to convey empathetic responses.
% Empathetic response generation has become a hot topic, 
% which requires fine-grained cognitive and emotional understanding as well as response expression.
% Existing large language models (LLMs) are strong in response expressions but weak in fine-grained cognitive and emotional understanding, i.e., reasoning of emotion cause words and fine-grained emotion detection. In contrast, small-scale empathetic models (SEMs) demonstrate complementary capabilities.
% To acquire their complementary advantages, we propose a Hybrid Empathetic Framework (HEF).
% HEF regards SEMs as flexible plugins to enhance LLMs from the fine-grained understanding of cognition and emotion.
% In the cognitive aspect, HEF employs an emotion cause perception strategy
% that guides LLMs to attend to emotion cause words emphasized by SEMs.
% This compensates for the deficiencies of LLMs in emotion cause reasoning, while attaining perceptual capabilities towards fine-grained emotion causes.
% In the emotional aspect, HEF adopts a two-stage emotion prediction strategy that directs LLMs to prioritize emotion categories deemed important by SEMs first, and then other categories later. This effectively alleviates the difficulties for LLMs in fine-grained emotion detection.
% Based on both strategies, LLMs accurately predict emotions and generate empathetic responses.
% We conduct experiments on the Empathetic-Dialogue dataset.
% The experimental results show that HEF effectively improves LLMs' fine-grained understanding while expressing more empathetic responses.
\end{abstract}

\section{Introduction}
As an important hot topic in dialogue tasks, empathetic response generation aims to finely understand the dialogue context from both emotional and cognitive perspectives, and express appropriate responses~\cite{rashkin2018towards,CEM2021,zhao2022EmpSOA,zhou2022case,yang2023exploiting}. Existing methods for empathetic response generation can be divided into small-scale empathetic models (SEMs) and large language models (LLMs).

% \begin{figure}
% \centering
% \includegraphics[width=78mm]{example_2.pdf}
% \caption{\label{fig example}
% An example of iteratively associating keywords to construe utterances.
% Words with the same color are associated terms.
% Associating and storing them in memory enables coherent and nuanced understanding of the content and feelings.
% }
% \end{figure}

\textbf{Small-scale Empathetic Models}. SEMs understand the dialogue context from an emotional or emotional-cognitive perspective and generate fitting responses~\cite{li2019empdg,li-etal-2022-kemp,CEM2021,majumder2020mime,lin2019moel,cai2023improving}.
SEMs have the capability of fine-grained understanding of dialogues, such as detecting fine-grained emotion categories across 32 classifications and  identifying the emotion causes behind them
~\cite{gao2021improving,Kim2021empathy},
but lack expressive capacities~\cite{bi2023diffusemp}.
% SEMs have the capability of fine-grained understanding of dialogues, such as 
% detecting fine-grained emotion categories across 32 classifications and reasoning the emotion causes behind them
% ~\cite{gao2021improving,Kim2021empathy}, but lack rich expressivity~\cite{bi2023diffusemp}.

\begin{figure}
\centering
\includegraphics[width=80mm]{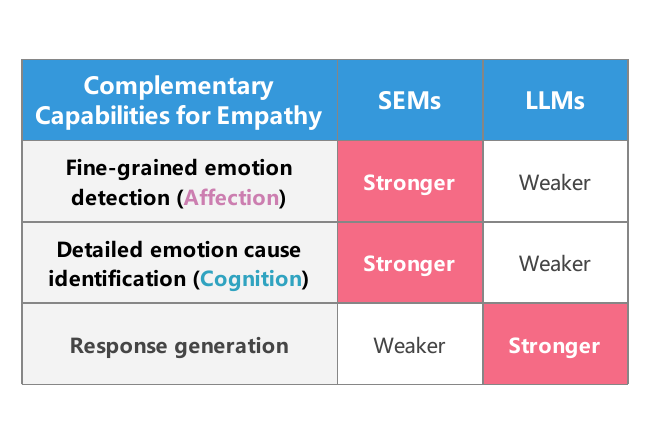}
\caption{\label{fig example}
An example illustration of complementary strengths between small-scale empathetic models (SEMs) and large language models (LLMs) for empathetic response generation.
}
\end{figure}

\textbf{Large Language Models}. 
LLMs have demonstrated superior performance on multiple tasks~\cite{qin2023chatgpt,wang2023robustness,chen2023llm,sun2023chatgpt}.
Despite this, LLMs have constraints on weight accessibility and computing resources.
To avoid such limitations, recent methods adopt non-finetuning approaches to validate emotional, cognitive, and expressive capabilities of LLMs on empathetic response generation.
These studies demonstrate that LLMs possess strong response expressions, yet lack fine-grained emotional and cognitive understanding capabilities essential for empathy~\cite{sorin2023large,zhao2023chatgpt}.
In terms of emotional capabilities, LLMs excel at coarse-grained emotion category detection, but underperform at fine-grained emotion prediction.
For example, LLMs achieve over 80\% accuracy on coarse-grained categories like 6 basic emotions~\cite{schaaff2023exploring}, but less than 40\% accuracy on fine-grained categories with 32 emotions~\cite{qian2023harnessing}.
In terms of cognitive capabilities, LLMs lack identifying abilities for detailed emotion causes, i.e., emotion cause words~\cite{yang2023towards}.
This inability leads to models failing to generate precise responses tailored to specific reasons.~\cite{Kim2021empathy}.

Overall, as shown in Figure \ref{fig example}, LLMs have stronger expressive capabilities but weaker fine-grained emotional and cognitive comprehension, while SEMs present complementary capabilities.
Therefore, how to combine the complementary capabilities of SEMs and LLMs to enhance empathy becomes an important problem.

To this end, we propose a Hybrid Empathetic Framework (HEF) for blending large language models and small-scale empathetic models to leverage their respective strengths.
HEF utilizes SEMs as flexible plugins in a non-finetuning way to enhance LLMs' emotional and cognitive capabilities.
Specifically, we enhance LLMs by constructing instructions from two aspects:
\textbf{Two-stage Emotion Prediction}.
We extract important emotion categories deemed most probable by SEMs, and guide LLMs to first infer emotions from these categories before considering other categories.
This sufficiently alleviates the difficulty for LLMs to predict fine-grained emotion categories, thereby enhancing the model's emotional capabilities.
\textbf{Emotion Cause Perception}.
We extract words emphasized by SEMs in the dialogue context as emotion causes and guide LLMs to attend to them at varied degrees.
This compensates for the cognitive deficiencies of LLMs in emotion cause identifying, while attaining perceptual capabilities towards detailed emotion causes.
Through the two strategies above, LLMs accurately understand fine-grained emotions and their subtle causes.
Based on the more accurate understanding, LLMs generate more tailored empathetic responses.

We conduct experiments on the Empathetic-Dialogue dataset~\cite{rashkin2018towards}.
The results show that HEF effectively improves LLMs' fine-grained emotional and cognitive understanding, while expressing proper empathetic responses.
%At the same time, we also analyze the characteristics of emotional causal words and two-stage emotions, which further demonstrate the effectiveness of HEF's in-depth reasons.
Overall, our contributions are as follows:

\begin{itemize}
\item 
%We find that the correlations between emotions and semantics promote accurate recognition of emotions and detection of important semantics, which has been overlooked in previous work.
We introduce a novel perspective of combining small-scale models with large language models for empathetic response generation.
\item 
We propose a new non-fine-tuning framework that effectively mitigates large language models' struggles in fine-grained emotional and cognitive understanding through a pluggable approach.

\item Experiments on the Empathetic-Dialogue dataset demonstrate the efficacy of the framework.
\end{itemize}

\section{Related Work}
Empathetic response generation aims to cognitively and emotionally understand the dialogue context and express appropriate responses~\cite{rashkin2018towards}.
Existing studies can be categorized into small-scale empathetic models (SEMs) and large language models (LLMs).

\textbf{Small-scale empathetic models}.
Small-scale empathetic models refer to models with relatively small parameters that are trained on specific datasets. 
SEMs can be divided into two lines.
The first is to understand emotions implied in the dialogues, including coarse-grained utterance-level emotions~\cite{rashkin2018towards,lin2019moel,majumder2020mime} and fine-grained word-level emotions~\cite{li2019empdg,li-etal-2022-kemp,gao2021improving, Kim2021empathy,yang2023exploiting}.
The second line enhances empathetic understanding through commonsense knowledge~\cite{CEM2021}, self-other awareness~\cite{zhao2022EmpSOA}, emotion-cognition alignment~\cite{zhou2022case}, dynamic commonsense fusion~\cite{cai2023improving}, and the multi-grained control diffusion framework~\cite{bi2023diffusemp}, given that empathy involves both emotional and cognitive aspects~\cite{davis1983measuring}.
Although these methods enhance empathy in various ways, their capabilities in response expression remain insufficient~\cite{bi2023diffusemp}.

\textbf{Large language models}.
Large language models have demonstrated exceptional performance on various tasks~\cite{qin2023chatgpt,wang2023robustness,chen2023llm,sun2023chatgpt}.
Due to the constraints on weight accessibility and computing resources of LLMs, non-fine-tuning approaches are adopted for empathetic response generation.
Existing studies evaluate LLMs' performance from various aspects.
~\citet{sorin2023large} and ~\citet{zhao2023chatgpt} demonstrate that LLMs possess strong capabilities in response expression.
~\citet{yang2023towards} argue that LLMs lack the cognitive understanding imperative for empathy, namely the reasoning of emotion cause words.
~\citet{schaaff2023exploring} and ~\citet{qian2023harnessing} show LLMs lack fine-grained emotional understanding abilities.
%Meanwhile, ~\citet{amin2023will} and ~\citet{yang2023towards} indicate LLMs underperform specialized small models in emotion perception.

Overall, SEMs have stronger fine-grained cognitive and emotional understanding but weaker response expression.
LLMs possess stronger response expression, but poorer fine-grained cognitive and emotional understanding.
That is, SEMs and LLMs present complementary capabilities.
To take full advantage of the strengths of SEMs and LLMs in empathetic response generation, we propose a hybrid framework (HEF) fusing both types of models. HEF incorporates SEMs as plugins to enhance LLMs' fine-grained understanding from perspectives of cognition and emotion.

\begin{figure*}
\centering
\includegraphics[width=140mm]{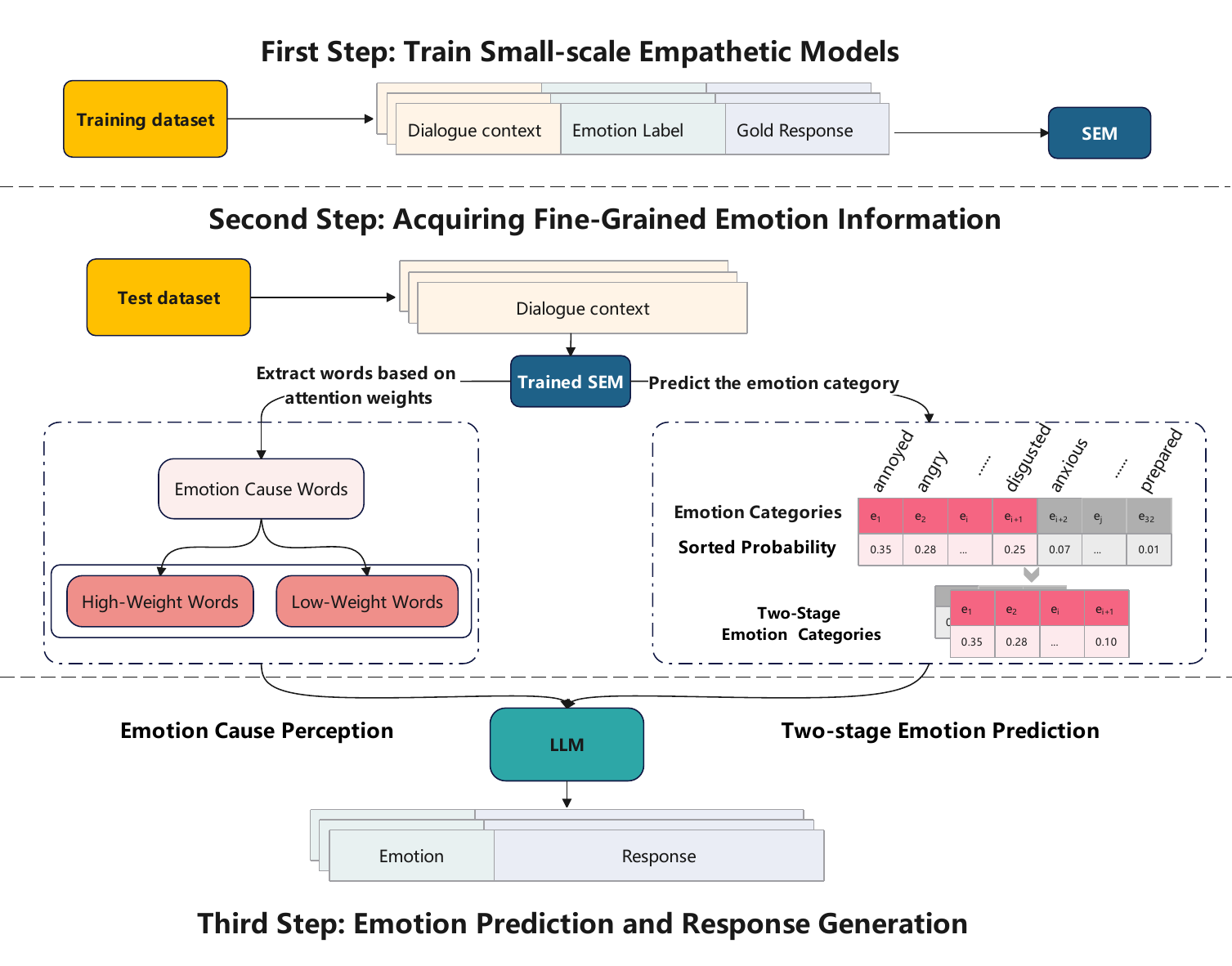}
\caption{\label{fig 2}
Overview of Hybrid Empathetic Framework (HEF).
}
\end{figure*}
\section{Method}
\subsection{Task Formulation}
The task of empathetic response generation is:
Given the context $D$ = [$U_1$, ..., $U_i$, ..., $U_{M}$] of a multi-turn dialogue, the model needs to predict the emotion $E$ of the dialogue and generate a response $Y$ = [$y_1$, $y_2$, ...,$y_j$, $y_N$] based on the predicted emotion. 
$U_i$ = [$w^i_1, w^i_2, ..., w^i_{m_i}$] represents the i-th utterance in the dialogue with $m_i$ words.
$E$ is a fine-grained emotion category, one of 32 emotions in our task.
$Y$ is the response with $N$ words.

\subsection{Overview}
As shown in Figure \ref{fig 2} and Algorithm 1, our proposed Hybrid Empathetic Framework (HEF) contains three main steps:
(1) Training Small-scale Empathetic Model (\textbf{Section \ref{Training Small-scale Model}}). We first train a small-scale empathetic model ESCM$_{tt}$\footnote{\url{https://github.com/wangyufeng-empty/TwoTree}} on the specific empathetic dataset, namely the EmpatheticDialogues dataset~\cite{rashkin2018towards}.
(2) Acquiring Fine-Grained Emotion Information (\textbf{Section \ref{Acquiring emotion cause Words}}). We then utilize the trained ESCM$_{tt}$ to acquire fine-grained emotion information, including emotion cause words and important emotion categories emphasized by ESCM$_{tt}$.

(3) Emotion Prediction and Response Generation (\textbf{Section \ref{Acquiring Important Emotion Categories}}).
Based on the acquired fine-grained emotion information, we leverage instructions to guide LLMs in predicting emotions and generating responses through emotion cause perception and a two-stage emotion prediction strategy.

\subsection{Training Small-scale Model}
\label{Training Small-scale Model}
The first step is to train a small-scale empathetic model ESCM$_{tt}$ on the Empathetic-Dialogue (ED) dataset.
Since ESCM$_{tt}$ has fewer parameters, it requires less computational resources and training time during the training process.
After training, compared to non-fine-tuned large language models such as ChatGPT, ESCM$_{tt}$ achieves higher accuracy in fine-grained emotion recognition on the ED dataset.
It is worth noting that, to demonstrate the efficacy of HEF, the emotion recognition capability of the chosen ESCM$_{tt}$ is not the optimal among small-scale empathetic models.
\subsection{Acquiring Fine-Grained Emotion Information}
The second step is to acquire fine-grained emotion information, including emotion cause words and important emotion categories.

\textbf{Acquiring Emotion Cause Words}.
\label{Acquiring emotion cause Words}
In classification models with attention mechanisms, the model tends to assign higher weights to words that contribute more to predicting the target class~\cite{yang2016hierarchical}.
Similarly, when predicting emotions, the ESCM$_{tt}$ tends to assign higher attention weights to words that contribute more to the emotions.
As with the previous method ~\cite{Kim2021empathy}, we treat these words as subtle causes of emotion prediction and refer to them as emotion cause words.
For each dialogue context in the test set $D_{T}$, 
we extract the top $k_1$ emotion cause words emphasized by ESCM$_{tt}$ and add them to the set $S$.
Then we compute the average emotion intensity~\cite{zhong2019knowledge, li-etal-2022-kemp} and average inverse document frequency (IDF) over words in set $S$.
In each dialogue context, words with both emotion intensity and IDF value greater than the average are defined as high-weight words, while the remaining context words existing in $S$ are defined as low-weight words.
By instructing the LLM to focus on and distinguish between high-weight words and low-weight words in the dialogue context, the model can more sensitively perceive subtle differences in emotional causes.
Meanwhile, we also instruct the LLM to jointly pay attention to correlations between high-weight words and low-weight words to understand emotion causes more comprehensively.

\textbf{Acquiring Important Emotion Categories}.
\label{Acquiring Important Emotion Categories}
Since the small-scale empathetic model ESCM$_{tt}$ has been trained on empathetic datasets, it thus has obvious advantages in understanding fine-grained emotional categories.
Table \ref{table escm_acc} shows the emotion accuracy of the ESCM$_{tt}$ on the Empathetic-Dialogue (ED) dataset.
We are surprised to find that ESCM$_{tt}$ achieves an 87\% accuracy in identifying the top 10 emotions.
Meanwhile, LLMs are weaker in recognizing fine-grained emotions but have higher accuracy in identifying coarse-grained emotions. For instance, ChatGPT's accuracy in classifying 32 emotion categories is less than 40\%, while its accuracy in classifying 6 emotion categories exceeds 80\%.
Consequently, we convert fine-grained emotion categories into coarse-grained ones, enabling large language models to prioritize the more probable coarse-grained emotional categories and subsequently infer emotions from the other categories. This strategy mitigates the issue large language models face in discerning fine-grained emotional categories.

Specifically, for each dialogue context, we first use ESCM$_{tt}$ to predict its emotion category.
We then rank these emotion categories in descending order by probability.
Next, we take the top $k_2$ emotions with the highest probabilities as the important emotion categories E$_{k_2}$.

\begin{table*}
\centering
\begin{tabular}{l}%p{15cm}
\hline
\textbf{Algorithm 1} Hybrid Empathetic Framework \\
\hline
\textbf{Require:} Test set $D_{T}$=\{$x_1$, ...,$x_i$,..., $x_{L}$\}, LLM M, small-scale empathetic model ESCM$_{tt}$, \\
\quad \quad \quad \quad empty set $S, S_{high}, S_{low}, S^e_{pred}$ \\
\textbf{Ensure:} Predicted emotion category E and generated response R \\
{\fontsize{8}{25}\selectfont \quad1:} \textbf{for} test sample $x_i$ \textbf{do} \\
{\fontsize{8}{25}\selectfont \quad2:} \quad Select top $k_{1}$ words with highest emotion attention weights in ESCM$_{tt}$ into set $S$ \\
{\fontsize{8}{25}\selectfont \quad3:} \quad Compute average emotion intensity $I^e_{avg}$ and average IDF $IDF_{avg}$ over words in $S$ \\
{\fontsize{8}{25}\selectfont \quad4:} \textbf{for} test sample $x_i$ \textbf{do} \\
{\fontsize{8}{25}\selectfont \quad5:} \quad \textbf{for} dialogue word $w_j$ in sample $x_i$ \textbf{do} \\
{\fontsize{8}{25}\selectfont \quad6:} \quad \quad If $w_j \in S$ \\
{\fontsize{8}{25}\selectfont \quad7:} \quad \quad \quad If $I^e_{w_j} > I^e_{avg}$ and $IDF_{w_j} > IDF_{avg}$ Add $w_j$ into set $S_{high}$\\
{\fontsize{8}{25}\selectfont \quad8:} \quad \quad \quad Else Add $w_j$ into set $S_{low}$ \\
{\fontsize{8}{25}\selectfont \quad9:} \quad \textbf{end for} \\
{\fontsize{8}{25}\selectfont \quad10:} \quad Select top $k_2$ emotions with highest probabilities $E^{e}_{pred}$ in ESCM$_{tt}$ into set $S^{e}_{pred}$ \\
{\fontsize{8}{25}\selectfont \quad11:} \textbf{end for} \\
{\fontsize{8}{25}\selectfont \quad12:} \textbf{for} test sample $x_i$ \textbf{do} \\
{\fontsize{8}{25}\selectfont \quad13:} \quad Construct instruction to:  \\
{\fontsize{8}{25}\selectfont \quad14:} \quad \quad Incorporate $w_{high}$ and $w_{low}$ to focus on emotional causal words \\
{\fontsize{8}{25}\selectfont \quad15:} \quad \quad Prioritize emotion categories in $E^{e}_{pred}$ first before considering other emotions.\\
{\fontsize{8}{25}\selectfont \quad16:} \quad Predict emotion category E and generate response R based on instruction \\
{\fontsize{8}{25}\selectfont \quad17:} \textbf{end for} \\
\hline
\end{tabular}
\label{table algorithm}
\end{table*}

\begin{table}
\centering
\begin{tabular}{ccccc}
\hline
\textbf{Models} & \textbf{Acc$_1$} & \textbf{Acc$_3$} & \textbf{Acc$_{10}$} & \textbf{Acc$_{20}$} \\
\hline
ESCM$_{tt}$ & 42.02 & 66.39 & 87 & 96.57 \\
\hline
\hline
\end{tabular}
\caption{\label{table escm_acc}
Emotion accuracy of the model, where Acc$_k$ represents the accuracy of the top $k$ predictions respectively.
}
\end{table}

\subsection{Emotion prediction and response generation}
The third step is to utilize LLMs to predict emotions and generate responses.

Based on the two types of fine-grained emotion information above, we construct an instruction.
The constructed instruction has two aspects:

\textbf{Emotion Cause Perception}.
We require LLMs to focus on the correlations between high-weight and low-weight words to gain a profound understanding of the subtle causes behind the dialogue.
Since the high-weight and low-weight words are divided into two different sets in the instruction, LLMs can also differentiate between them.
Specifically, the constructed emotion cause words inevitably contain noise. 
For LLMs with weaker understanding abilities except ChatGPT, we do not consider this strategy.

\textbf{Two-stage Emotion Prediction}.
We require LLMs to prioritize emotions in the important emotion categories focused on by ESCM$_{tt}$ when predicting emotions, and then consider other emotions.

By inputting the constructed instruction into LLMs, the model predicts the possible emotions $E$ of the dialogue.

\textbf{Response Generation}.
LLMs generate appropriate responses after carefully considering the dialogue context and the predicted emotions $E$.  
It is noteworthy that the emotion cause perception, two-stage emotion prediction, and response generation are different logical parts of the same prediction process.

\subsection{Baselines}
To validate the effectiveness of HEF, we select the following state-of-the-art (SOTA) small-scale empathetic models and large language models:

\textbf{Small-Scale Empathetic Models}.
\textbf{KEMP}~\cite{li-etal-2022-kemp} captures the implicit knowledge implied in dialogues through ConceptNet~\cite{speer2017conceptnet} to enhance emotion understanding.
\textbf{CEM}~\cite{CEM2021} introduces COMET reasoning knowledge~\cite{hwang2021comet}, providing a more comprehensive understanding of empathy from both emotional and cognitive perspectives.
\textbf{CASE}~\cite{zhou2022case} aligns emotions and cognition from both coarse-grained and fine-grained aspects to enhance empathy.
\textbf{ESCM}~\cite{yang2023exploiting} utilizes dynamic emotion-semantic correlation to improve the model's emotional understanding.
\textbf{ESCM$_{tt}$} is an improved version of ESCM, focusing on the dynamic emotion-semantic correlation from both coarse-grained and fine-grained perspectives.

\textbf{Large Language Models}. \textbf{Llama2$_{7b}$} and \textbf{Llama2$_{13b}$}~\cite{touvron2023llama} are large language models developed by Meta AI, with 7 billion and 13 billion parameters respectively.
\textbf{ChatGLM3$_{6b}$}~\cite{zeng2022glm,du2022glm} is a Chinese-English hybrid open-source large language model jointly released by Zhipu AI and the KEG laboratory at Tsinghua University.
\textbf{Mistral$_{7b}$}~\cite{jiang2023mistral} is an open-source large language model with 7.3 billion parameters created by Mistral AI.
\textbf{ChatGPT} is a large language model developed by Open AI, with excellent cognitive understanding and response expression capabilities.

\subsection{Implementation Details}
We conduct experiments on the Empathetic-Dialogue dataset~\cite{rashkin2018towards}, which is a dialogue dataset with 32 fine-grained emotion categories.  
For the small-scale empathetic model ESCM$_{tt}$, we retain all parameters of the original model.  
Meanwhile, we set the number of emotion cause words $k_1$ to 1. 
The number of most important emotion categories $k_2$ is set to different optimal values for different LLMs due to their diverse characteristics.
As for LLMs, we choose Llama2$_{7b}$, Llama2$_{13b}$, ChatGLM3$_{6b}$, Mistral$_{7b}$, and ChatGPT as the large language models for HEF.
We experiment with the ChatGPT model through the API interface, while other models primarily experiment using the LLaMA-Factory framework\footnote{\url{https://github.com/hiyouga/LLaMA-Factory}} on the NVIDIA RTX 3090 GPU.
Furthermore, to validate the model's performance, we use GPT4.0 for human-like evaluation.

\begin{table*}
\centering
\begin{tabular}{cccccc}
\hline
\textbf{Model Type} & \textbf{Models} & \textbf{Accuracy} $\uparrow$ & \textbf{Perplexity} $\downarrow$ & \textbf{Distinct-1} $\uparrow$ & \textbf{Distinct-2} $\uparrow$ \\
\hline
\multirow{5}{*}{\centering \shortstack{Small-Scale \\Empathetic Models \\(SEMs)}}
& KEMP & 39.31 & 36.89 & 0.55 & 2.29 \\ % & 3.49 & 3.92 & 3.65\\
& CEM & 39.11 & 36.11 & 0.66 & 2.99 \\ % & 0.0 & 0.0 & 0.0\\
& CASE & 40.2 & 35.37 & 0.74 & 4.01 \\ 
& ESCM & 41.19 & 34.82 & 1.19 & 4.11 \\
& ESCM$_{tt}$ & 42.02 & 35.07 & 1.39 & 4.42 \\
%EmpSOA & 48.32 & 35.02 & 0.71 & 3.96 \\
\hline
\multirow{5}{*}{\centering \shortstack{Large \\Language Models \\(LLMs)}}
& Llama2$_{7b}$ & 3.06 & - & 26.18 & 66.93 \\
& Llama2$_{13b}$ & 4.52 & - & 5.46 & 29.17 \\
& ChatGLM3$_{6b}$ & 24.31 & - & 37.75 & 75.03 \\
& Mistral$_{7b}$ & 26.77 & - & 3.76 & 23.85 \\
& ChatGPT & 37.9 & - & 3.58 & 21.38 \\
\hline
\multirow{5}{*}{\centering \shortstack{HEF-based Models \\(Ours)}}
& Llama2$^{c_{10}}_{7b}$ & 5.57 & - & 24.02 & 66.37 \\
& Llama2$^{c_3}_{13b}$ & 7.09 & - & 6.24 & 31.86 \\
& ChatGLM3$^{c_3}_{6b}$ & 27.21 & - & \textbf{42.23} & \textbf{80.08} \\
& Mistral$^{c_3}_{7b}$ & 31.36 & - & 3.41 & 22.69 \\
& ChatGPT$^{c_{20},w_{1}}$ & \textbf{45.63} & - & 3.36 & 20.9 \\
\hline
\end{tabular}
\caption{\label{table automatic}
Results of automatic evaluation, 
where models with the superscript $w_i$ employ the emotion cause perception strategy, and those with the trademark $c_j$ employ the two-stage emotion prediction strategy. $i$ and $j$ are the number of emotion cause words and the number of important emotion categories, respectively.
}
\end{table*}

\subsection{Evaluation Metrics}
To validate the effectiveness of the Hybrid Empathy Framework (HEF), we employ the following two evaluation methods: 

\textbf{Automatic Evaluation}.
Following previous methods ~\cite{li-etal-2022-kemp,CEM2021}, we employ perplexity, accuracy, Distinct-1, and Distinct-2 ~\cite{li2015diversity}.
Perplexity reflects the fluency of the responses, with lower scores indicating better performance. However, perplexity does not apply to large language models due to the differences in their vocabularies~\cite{qian2023harnessing}.
Accuracy measures the model's emotion perception capability. The stronger the emotion perception ability, the higher the score.
Distinct-1 and Distinct-2 evaluate the diversity of responses at the unigram and bigram levels, respectively. 
For small-scale models, the higher the diversity score, the richer the information reflected.
Whereas for large language models, we find that to a certain extent, the lower the diversity, the higher the quality of the responses.
It is worth noting that, as BLEU~\cite{papineni2002bleu} does not apply to the empathetic response generation task~\cite{liu2016not, CEM2021}, we do not consider this metric.

\begin{table}
\centering
\begin{tabular}{cccc}
\hline
\textbf{Comparisons} & \textbf{Aspects} & \textbf{Win} & \textbf{Lose}\\
\hline
\multirow{3}{*}{\centering \shortstack{ChatGPT$^{c_{20},w_1}$ \\ vs. ChatGPT}} & Emp. & \textbf{86} & 1 \\
& Rel. & \textbf{44} & 0 \\
& Flu. & \textbf{32} & 0 \\
\hline
\multirow{3}{*}{\centering \shortstack{Mistral$^{c_3}_{7b}$ \\ vs. Mistral$_{7b}$}} & Emp. & \textbf{48} & 40 \\
& Rel. & \textbf{51} & 23\\
& Flu. & \textbf{34} & 21\\
\hline
\multirow{3}{*}{\centering \shortstack{ChatGLM$3^{c_3}_{6b}$ \\ vs. ChatGLM3}}  & Emp. & \textbf{63} & 22 \\
& Rel. & \textbf{54} & 16\\
& Flu. & \textbf{52} & 7 \\
\hline
\end{tabular}
\caption{\label{table human}
Results of human-like evaluation.
}
\end{table}
\textbf{Human-like Evaluation Metrics}.
Since the evaluation based on GPT4 is highly consistent with human evaluation~\cite{qian2023harnessing}, we employ GPT4 to replace the time-consuming manual evaluation.
Following previous methods~\cite{li-etal-2022-kemp,yang2023exploiting}, we use an A/B test to compare the baselines and HEF-Based models.
We first randomly select 100 dialogue samples and pairwise compare the effects of the baseline and HEF-based models.
For the same dialogue, if the HEF-based model performs better, we increment the score for \emph{Win}.
If the HEF-based model performs worse, we increment the score for \emph{Lose}.
To comprehensively evaluate the model's performance, we assess it from the perspectives of Empathy (Emp.), Relevance (Rel.), and Fluency (Flu.).
Empathy measures whether the emotional response is appropriate.
Relevance measures whether the response is relevant to the content and topic of the dialogue context.
Fluency measures whether the response is natural, fluent, and aligns with human expression habits.

\section{Results and Analysis}
\subsection{Main Results}
% \begin{table}
% \centering
% \begin{tabular}{cccccc}
% \hline
% \textbf{Models} %& \textbf{Ablated module } 
% & \textbf{Acc} & \textbf{PPL} & \textbf{Dist-1} & \textbf{Dist-2}\\
% \hline
% IAMM & \textbf{55.92} & 35.66 & \textbf{2.09} & \textbf{7.03} \\

% \hline
% w/o EA
% & 51.34 & 34.67 & 1.01 & 3.45\\

% w/o IA
% & 51.37 & 34.59 & 0.92 & 3.02\\

% w/o WS
% &  &  &  & \\

% \hline
% \end{tabular}
% \caption{\label{table ablation}
% Results of the ablation experiments.
% }
% \end{table}

\textbf{Automatic Evaluation Results}.
The results of the automatic evaluation metrics are shown in Table \ref{table automatic}. 
The results indicate that SEMs and LLMs have complementary strengths in understanding and expression.
That is, SEMs demonstrate better fine-grained emotion comprehension abilities, while LLMs exhibit better expression capabilities.
Additionally, the HEF-based model outperforms both SEMs and LLMs in terms of comprehension and expression capabilities.

In terms of emotion accuracy, the HEF-based model outperforms SEMs and LLMs. 
This is primarily because HEF-based models have higher accuracy in predicting coarse-grained emotion categories (e.g. 6 classes), while the two-stage emotion strategy converts the 32 emotion classification into a coarse-grained emotion classification task, such as 3 categories.
This enhances the emotion classification accuracy of the HEF-based model.
Additionally, we find Llama2$_{7b}$ and Llama2$_{13b}$ perform significantly worse than ChatGLM3$_{6b}$ and Mistra$_{7b}$.
This is because Llama2$_{7b}$ and Llama2$_{13b}$ have relatively poor instruction-following abilities without fine-tuning, resulting in predicted emotions not belonging to the 32 emotion categories.

In terms of diversity, the HEF-based model outperforms SEMs, demonstrating the HEF-based model's superior expression capabilities.
However, the HEF-based model underperforms LLMs regarding diversity.
Simultaneously, ChatGPT, with stronger expression capabilities, also shows lower diversity compared to other HEF-based models.
At the same time, previous studies ~\cite{ayers2023comparing,sorin2023large} have also shown that the quality of lengthy and complex responses is likely to be inferior to succinct ones.
Based on the above experimental results, we speculate that the LLMs' understanding of the information is more accurate, the expressed responses are more precise and concise. 
Thus, the relatively lower diversity to some extent indicates stronger understanding and expression abilities of the LLMs.

%This primarily results from the model generating informative and relevant responses based on essential correlated words and accurate emotional delivery.

\textbf{Human-like Evaluation Results}.
Table \ref{table human} shows the performance of the three strongest models on human-like metrics.
The HEF-based models demonstrate better empathy than the baselines, primarily due to the two-stage emotion prediction strategy, which facilitates accurate emotion understanding.
The advantage in relevance stems mainly from the emotion cause perception strategy that captures important emotion cause words. The models express more pertinent responses through these important words.
The fluency advantage is due to both strategies promoting more natural response formulation in terms of emotion and wording.

\begin{table}
\centering
\begin{tabular}{cccc}
\hline
\textbf{Models} & \textbf{Acc} & \textbf{Dist-1} & \textbf{Dist-2} \\
\hline
ChatGPT & 37.9 & 3.58 & 21.38 \\
ChatGPT$^{c_{20},w_{1}}$ & \textbf{45.63} & 3.36 & 20.9 \\
\hline
ChatGPT$^{c_{20}}$ & 45.44 & \textbf{3.59} & 21.29 \\
ChatGPT$^{w_{1}}$ & 38.66 & 3.57 & \textbf{21.41} \\
\hline
\end{tabular}
\caption{\label{table automatic ablation}
Results of automatic evaluation for ablation study.
}
\end{table}

\subsection{Ablation Studies}
To further validate the effectiveness of HEF, we construct the following ablation models:
(1) \textbf{ChatGPT$^{c_{20}}$} is the model that only employs two-stage emotion prediction.
(2) \textbf{ChatGPT$^{w_{1}}$} is the model that only employs emotion cause preception.
Note that other LLMs lack strong understanding capabilities and cannot comprehend emotion cause words with noise.
Therefore, we only have ChatGPT, with its excellent understanding capabilities, focus on emotion cause words with noise.
For this reason, we conduct ablation experiments solely on ChatGPT.

\begin{table}
\centering
\begin{tabular}{cccc}
\hline
\textbf{Comparisons} & \textbf{Aspects} & \textbf{Win} & \textbf{Lose}\\
\hline
\multirow{3}{*}{\centering \shortstack{ChatGPT$^{c_{20},w_1}$ \\ vs. ChatGPT$^{w_1}$}} & Emp. & \textbf{89} & 3 \\
& Rel. & \textbf{68} & 2 \\
& Flu. & \textbf{39} & 0 \\
\hline
\multirow{3}{*}{\centering \shortstack{ChatGPT$^{c_{20},w_1}$ \\ vs. ChatGPT$^{c_{20}}$}} & Emp. & \textbf{77} & 7 \\
& Rel. & \textbf{74} & 0 \\
& Flu. & \textbf{65} & 0 \\
\hline
\end{tabular}
\caption{\label{table human ablation}
Results of human-like evaluation for ablation study.
}
\end{table}

Tables \ref{table automatic ablation} and \ref{table human ablation} show the results of ablation models on automatic and human-like metrics, respectively.
The automatic evaluation results indicate that the emotion cause perception strategy improves response expression, while the two-stage emotion prediction enhances emotion understanding.
The human-like evaluation results suggest that both strategies contribute to empathy, relevance, and fluency.
Emotion cause perception mainly contributes to relevance and fluency, whereas two-stage emotion prediction contributes more to empathetic responses.
%Overall, the experimental results reaffirm the validity of both strategies.

\subsection{Hyperparameter Experiments}
To validate the impact of different hyperparameters on the model, we conduct the following parameter experiments.

\textbf{Number of Emotion Cause Words}.
We conduct experiments on the model ChatGPT$^{w_{k_1}}$ based on the emotion cause perception strategy, where $k_1$ is the number of emotion cause words.
The results in Table \ref{table hyperparameter} show that as $k_1$ increases, emotion accuracy continuously decreases while response diversity keeps increasing.  
This is mainly because as the number of emotion cause words increases, so does the noise.
The increased noise affects accurate emotion identification and precise response expression.

\begin{table}
\centering
\begin{tabular}{ccccc}
\hline
\textbf{Metrics} & \textbf{$k_1@1$} & \textbf{$k_1@5$} & \textbf{$k_1@10$} & \textbf{$k_1@15$}\\
\hline
Accuracy & 38.66 & 38.32 & 38.17 & 37.84\\
Distinct-1 & 3.57 & 3.54 & 3.61 & 3.63\\
Distinct-2 & 21.41 & 21.93 & 22.11 & 22.11\\
\hline
\end{tabular}
\caption{\label{table hyperparameter}
Performance of ChatGPT$^{w_{k_1}}$ with varying numbers of emotion cause words.
}
\end{table}

\begin{figure}
\centering
\includegraphics[width=70mm]{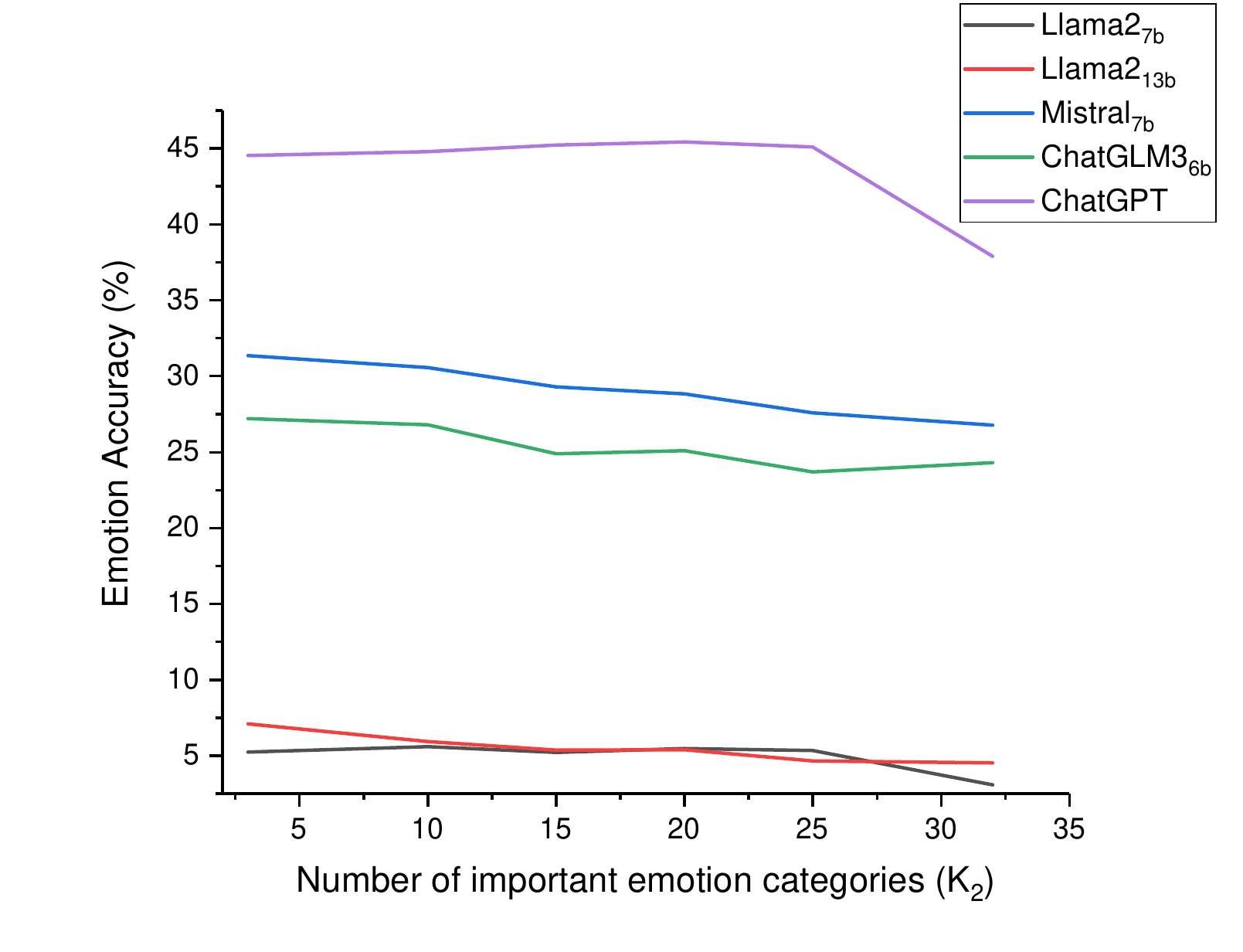}
\caption{
Emotion accuracy across different models.
\label{fig acc ablation}
}
\end{figure}

\textbf{Number of Important Emotion Categories}.
We validate the impact of varying numbers of important emotion categories $k_2$ on emotion accuracy.
The experimental results are shown in Figure \ref{fig acc ablation}.
The results indicate differences in the optimal number of emotion categories for different language models, primarily attributed to discrepancies in language understanding capabilities.

\subsection{Case Study}
To verify the effectiveness of HEF, we conduct case studies.
The details are shown in Appendix \ref{Appendix A}.

\section{Conclusion and Future Work}
In this paper, we have proposed a Hybrid Empathetic Framework (HEF) for empathetic response generation.
HEF treats small empathetic models (SEMs) as plugins to compensate for the deficiency of large language models (LLMs) in fine-grained emotional and cognitive understanding,
utilizing two strategies: two-stage emotion prediction and emotion cause perception.
The two-stage emotion prediction strategy alleviates the difficulty of LLMs in detecting fine-grained emotion categories by prioritizing the important emotion categories emphasized by SEMs.
The emotion cause perception strategy addresses the issue of LLMs' inadequate identification of detailed emotion causes by attending to key emotion words emphasized by SEMs, leveraging the key emotion words that SEMs attend to.
Our experiments demonstrate that HEF enhances LLMs' fine-grained cognitive and emotional understanding and generates more empathetic responses.

In the future, we will further explore the effectiveness of HEF on more tasks, as this framework has low dependence on models and tasks.
Meanwhile, we will explore more evaluation metrics for LLMs on empathetic response generation.

\section{Limitations}
Our work has the following limitations:
(1) We have only validated the effectiveness of HEF on the empathetic response generation task.
This method is also applicable to other tasks, especially multi-classification tasks. 
In the future, we will validate the effectiveness of this method on more tasks.
(2) Since LLMs possess stronger cognitive understanding and expression capabilities, the evaluation metrics used for SEMs are no longer applicable.
The metrics we employed cannot comprehensively evaluate the various capabilities of LLMs. 
Therefore, we will explore more suitable evaluation metrics in the future.

\section{Ethical Considerations}
We use the publicly available Empathetic-Dialogue dataset, which does not contain any information that involves ethical risks.
We adhere to the relevant guidelines when utilizing ChatGPT and GPT4.0.
Additionally, other models mentioned in the paper are open-source, and we have used these models in compliance with their respective guidelines.

%\clearpage

\bibliography{custom}
\bibliographystyle{acl_natbib}

 \appendix
 \section{Appendix}
 \label{Appendix A}
 As shown in Table \ref{table case}, we present the responses generated by the HEF-based model.

In the first case, ChatGPT generates a fluent response, but the sentence lacks expressions with delicate and highly relevant words.
ChatGPT$^{c_{20}, w_1}$ accurately understands the emotion of ``prepare'' expressed by the speaker, and clearly expresses the emotion ``wonderful''.
At the same time, ChatGPT$^{c_{20}, w_1}$ also expresses more words with strong relevance according to the dialogue content, such as ``emergency'', ``financial cushion''.

In the second case, ChatGPT understands the negative emotion ``sad'' expressed by the speaker. Due to the lack of in-depth understanding, ChatGPT only gave a superficial response.
ChatGPT$^{c_{20}, w_1}$ understands the negative emotion in the dialogue, and realizes that the speaker needs more comfort based on the emotion cause words ``passed away''.
Therefore, ChatGPT$^{c_{20}, w_1}$ expresses a more empathetic response of "I am here for you".

\begin{table*}[h!]
\centering
\begin{tabular}{p{2.3cm}|p{13cm}}
%\begin{tabular}{ll}

\hline
\textbf{Emotion} & \textbf{Proud}\\
%\textbf{Situation} & I was so \textcolor{c7}{excited} of \textcolor{green}{my daughter} when \textcolor{green}{she} was \textcolor{c9}{accepted into harvard}. \textcolor{green}{She} worked so hard through high school and has \textcolor{c9}{achieved her dream}.\\
\textbf{Speaker} & So I had \textcolor{green}{an emergency come up} for my \textcolor{blue}{family} and I had enough \textcolor{blue_deep}{money} to cover it. I feel great for \textcolor{c8}{preparing}!\\
\textbf{Listener} & That is always a good feeling. such a \textcolor{c7}{relief}.\\
\textbf{Speaker} & \textcolor{c9}{Definitely}! And they are paying me back, so it is not like I am losing \textcolor{blue_deep}{money} either.\\
\textbf{Gold} & It was nice of you to step up and help out.\\
\hline
\textbf{ChatGPT} & I'm so \textcolor{c9}{glad to} hear that everything worked out for you. It's \textcolor{c9}{wonderful} that you were able to help your \textcolor{blue}{family} in their time of need.\\
% \hdashline
% \textbf{ChatGPT$^{c_{20}}$} & That's \textcolor{c9}{wonderful} to hear that you were \textcolor{c8}{prepared} for the \textcolor{green}{emergency}. It must bring you a sense of \textcolor{c7}{relief} and \textcolor{c9}{hope} for the future.\\
% \hdashline
% \textbf{ChatGPT$^{w_{1}}$} & It's impressive that you were able to \textcolor{c8}{prepare} for the \textcolor{green}{emergency}. It's always a \textcolor{c7}{relief} to be ready for \textcolor{green}{unexpected situations}.\\
\hdashline
\textbf{ChatGPT$^{c_{20},w_{1}}$} & That's \textcolor{c9}{wonderful} to hear that you were \textcolor{c8}{prepared} for the \textcolor{green}{emergency} and that you'll be getting \textcolor{blue_deep}{reimbursed}. It's \textcolor{c9}{definitely} a \textcolor{c7}{relief} to have that \textcolor{blue_deep}{financial cushion}.\\
\hline
\textbf{Emotion} & \textbf{Sentimental}\\
%\textbf{Situation} & I was so \textcolor{c7}{excited} of \textcolor{green}{my daughter} when \textcolor{green}{she} was \textcolor{c9}{accepted into harvard}. \textcolor{green}{She} worked so hard through high school and has \textcolor{c9}{achieved her dream}.\\
\textbf{Speaker} & I always feel \textcolor{c9}{sad} when I think about my dad, he \textcolor{blue}{passed away} few years back.\\
\textbf{Gold} & Oh god .. So sorry to hear that. May I ask how did it happen?\\
\hline
\textbf{ChatGPT} & I'm so sorry to hear about your dad. Losing a loved one is \textcolor{blue}{incredibly tough} and it's normal to still feel \textcolor{c9}{sad} years later.\\
%\hdashline
%\textbf{ChatGPT$^{c_{20}}$} & I understand. Losing a loved one can be incredibly difficult, and it's natural to feel sad when thinking about them.\\
%\hdashline
%\textbf{ChatGPT$^{w_{1}}$} & I'm sorry to hear about your loss. It's understandable to feel sad when thinking about your dad. Do you want to share a happy memory of him?\\
\hdashline
\textbf{ChatGPT$^{c_{20},w_{1}}$} & I'm so sorry to hear about your dad. It's \textcolor{c9}{completely} normal to feel \textcolor{c9}{sad} when you think about him. If you ever need to talk, \textcolor{blue}{I'm here for you}.\\
\hline
\end{tabular}
\caption{\label{table case}
Case Study of HEF-based models and Benchmarks, in which color-coded words have related semantics or emotions.
}
\end{table*}

\end{document}